\documentclass[aps,amsmath,amssymb,reprint,longbibliography]{revtex4-1}
\usepackage[utf8]{inputenc}
\usepackage{bm}
\usepackage{graphicx}
\usepackage[unicode=true,colorlinks=true,urlcolor=blue,citecolor=blue]{hyperref}
\usepackage{natbib}

\begin{document}

\title{Effect of compensation on magnetic ordering in (Ga,Mn)As}

\author{I.A. Kokurin}
\email[E-mail:]{ivan.a.kokurin@gmail.ru} \affiliation{Ioffe
Institute, 194021 St. Petersburg, Russia}

\author{N.S. Averkiev}
\affiliation{Ioffe Institute, 194021 St. Petersburg, Russia}

\begin{abstract}
The real mechanism of ferromagnetism in III-V diluted magnetic
semiconductors such as Ga$_{1-x}$Mn$_x$As is still debated. The
experimental results point out the realization of appreciable
ferromagnetism at filling of the impurity band close to one-half
that can be expressed in terms of compensation degree. Based on our
recent results on the existence of two competing mechanisms of pair
indirect exchange (with ferro- and antiferro alignment of Mn $d$
shells), a theory of magnetic ordering in (Ga,Mn)As is presented.
The molecular-field-like equation is derived in which the effective
exchange constant contains all necessary information about material,
including the Mn content, the compensation degree, the crystal
coordination, and the parameters of two competing pair exchange
interactions. The Curie temperature depends crucially on the
compensation degree. Depending on Mn content and compensation
degree, three possible magnetic phases (paramagnetic, ferromagnetic,
and spin-glass-like) are predicted. However, the spin-glass-like
phase in (Ga,Mn)As is apparently not realized due to the additional
inevitable compensation caused by As antisites. The phase transition
into the ferromagnetic state occurs not only with Mn content
variation but through the change of the compensation degree as well.
The new mechanism of photoferromagnetic effect is predicted and
discussed in the context of possible clarification of the magnetic
ordering mechanism. It is assumed that the main obstacle to a
possible increase in the Curie temperature is the increasing
contribution of the antiferromagnetic superexchange between the
nearest-neighbor Mn's, rather than compensatory (electric or
magnetic) effects from Mn interstitials, as was previously stated.
The comparison of our predictions with available experimental data
is presented.
\end{abstract}

\date{\today}

\maketitle

\section{Introduction}
\label{Intro}

Thirty years ago the technology of low-temperature molecular beam
epitaxy was successfully applied for the growth of (Ga,Mn)As
\cite{Ohno1996}. This member of the diluted magnetic semiconductor
(DMS) family still attracts attention due to its potential
application in spintronics and ability to be integrated with
traditional semiconductor electronics.

The design of DMS materials with high Curie temperature remains a
challenge. The growth of DMS samples with given properties must be
associated with an understanding of the processes occurring in them.
However, the mechanism of ferromagnetism in Ga$_{1-x}$Mn$_x$As is
still debated. There is uncertainty relative to which kind of holes
mediates the exchange: valence band (VB) holes or holes of
Mn-derived impurity band (IB).

In the first decade of (Ga,Mn)As existence, the ferromagnetism was
explained in the spirit of Zener or Ruderman-Kittel-Kasuya-Yosida
(RKKY) models~\cite{Dietl2000,Dietl2001,Jungwirth2005}, i.e., the
ferromagnetism was claimed to be mediated by mobile VB holes. The
Zener-type theory implies the possibility of increasing the Curie
temperature $T_C$ due to an increase in the hole concentration $p$
and the manganese content $x$. However, the room-temperature limit
for (Ga,Mn)As is still not surpassed. This suggests that it is time
to rethink our understanding of magnetism in (Ga,Mn)As.

In recent years, a large number of arguments have been obtained
indicating that IB holes, rather then VB holes, are responsible for
ferromagnetism in
(Ga,Mn)As~\cite{Ohya2011,Dobrowolska2012,Muneta2013,Furdyna2023}.
The IB-based ferromagnetism in (Ga,Mn)As has been modeled within the
mean-field \cite{Berciu2001,Bhatt2002} and Monte Carlo simulations
\cite{Mayr2002,Kennett2002}. It has been shown that the spatial
inhomogeneity in the magnetic subsystem enhances the ferromagnetic
transition temperature \cite{Berciu2001}. However, some researchers
still defend the original point of view regarding the involvement of
VB holes. This confrontation is known as the \textit{battle of the
bands}~\cite{Samarth2012}.

In the zinc-blende lattice of III-V semiconductors, such as GaAs, Mn
can manifest itself in different ways: as a magnetic acceptor ${\rm
Mn_{Ga}}$ and as an interstitial double-charged donor ${\rm
Mn_I}$~\cite{Yu2008}. The latter provide electrical compensation,
which means the presence of two type of acceptors (neutral and
ionized). The interaction between different states of magnetic
impurities leads to intriguing consequences. In particular, the
all-optical manipulation of electron magnetization based on the
above-mentioned interactions has been predicted~\cite{Kokurin2020}.

Besides charge compensation ${\rm Mn_I}$'s reduce an amount of
magnetic moments participating in ferromagnetism (magnetic
compensation). This is due to Coulomb attraction of ionized ${\rm
Mn_I^{2+}}$ donors and ionized ${\rm Mn_{Ga}^-}$ acceptors with the
following antiferromagnetic (AFM) alignment of $d$
shells~\cite{Blinowski2003}. Thus, the amount of substitutional Mn's
$x_s$ participating in ferromagnetism is reduced by the amount $x_I$
of Mn interstitials, $x_{eff}=x_s-x_I$.

The magnetic properties are determined by the Fermi-level position
inside the IB (or, in other words, by the compensation degree). In
undoped (Ga,Mn)As samples, compensation is usually provided by ${\rm
Mn_I}$ interstitials, i.e., this is self-compensation. The excess
amount of Mn interstitials can be eliminated by means of
low-temperature annealing \cite{Potashnik2001} that leads to
enhancement of remanent magnetization and the Curie temperature. It
should be mentioned that the Fermi level position is determined by
certain feedback. An increase of ${\rm Mn_{Ga}}$ content lowers the
Fermi level and leads to the appearance of ${\rm Mn_I}$
interstitials, which in turn compensate ${\rm Mn_{Ga}}$ acceptors
and elevate the Fermi level giving an opportunity to introduce Mn in
the cation sublattice again. However, the magnetic compensation
excludes a significant portion of manganese from a participation in
ferromagnetism. From this point of view, an additional doping with
nonmagnetic donors seems more appropriate, since it provides the
necessary electrical compensation and can reduce (or almost exclude)
the Mn interstitial appearance. Finally, this increases the amount
of magnetic moments involved in ferromagnetism.

An additional $p$-type doping by nonmagnetic
acceptors~\cite{Wojtowicz2004} (including modulation
doping~\cite{Furdyna2004}) allows one to control the compensation
degree (CD) and in turn the Curie temperature. The Fermi-level
position depends on the method of introducing acceptors: either into
a layer grown before the growth of the magnetic layer or into a
layer grown after the magnetic one. The $n$-type counterdoping
\cite{Cho2008a} allows one to diminish the amount of interstitials
by means of the Fermi level control as well. Eventually, the
$n$-type counterdoping by nonmagnetic donors in the nonmagnetic
layer, grown before the magnetic one, can completely exclude
magnetic compensation and in turn maximize $x_{eff}$.

In our recent work \cite{Kokurin2024} the indirect exchange
interaction was considered to be mediated by Mn-acceptor localized
holes, i.e., the magnetism is associated with IB. Two competing
types of indirect exchange with ferromagnetic (FM) and AFM alignment
of $d$-shell spins were found. The predominance of one of them
depends on the CD $K$ or position of the Fermi level within the IB.
The predominant FM pair alignment is realized at $K\sim 1/2$
\cite{Kokurin2024} in agreement with IB half
filling~\cite{Dobrowolska2012}.

In this work the theory of FM order in (Ga,Mn)As is presented. The
clarification of the compensation (self-compensation) role is the
key point. The temperature dependence of a remanent magnetization
and the dependence of Curie temperature $T_C$ on the Mn content and
CD are studied. The usual thermodynamic approach (see, for instance,
Ref.~\cite{Kubo1988}) is adopted for the case of DMS. The main
differences are the following: (i) the necessity to take into
account an interaction between magnetic centers placed relative to
each other in distant coordination spheres (CSs) (the
nearest-neighbor approach is wrong); (ii) the disordered structure
(solid solution/alloy) of DMS; (iii) the presence of two competing
types of pair exchange interactions.

The paper is organized as follows. Section~\ref{Theory} contains the
derivation of equations describing the remanent magnetization and
Curie temperature in the disordered (Ga,Mn)As alloy. The analysis of
derived equations is presented in Sec.~\ref{Discussion}. The
interpretation of some experimental data in terms of our theory is
discussed in Sec.~\ref{Compare}. A brief summary of the results and
further prospects are presented in Sec.~\ref{Conclusion}. Some
details of experimental data processing are given in the Appendix.

\section{Derivation of molecular-field-like equation}
\label{Theory}

Let us employ a consistent statistical approach for consideration of
the ferromagnetism in DMS. The magnetization considered corresponds
to the alignment of the magnetic moments of the $d$ shells of Mn
ions, while, as is often the case, the contribution to the
magnetization of charge carriers (holes) is considered to be
negligible (see, for instance, Ref.~\cite{Berciu2001}). Here the
compensation by nonmagnetic donors is considered; i.e., all Mn's are
in substitutional positions (${\rm Mn^0_{Ga}}$ or ${\rm Mn^-_{Ga}}$)
only. This allows one to use the concentration $x$ of the magnetic
component and CD $K$ as independent variables. The self-compensation
by interstitial ${\rm Mn_I}$ donors can be simply introduced into
the problem. The macroscopic volume is considered with $N$ sites of
the cation sublattice. The number of lattice sites substituted by Mn
is equal to $xN$. The CD value of $K$ ($0\leq K\leq 1$) means that
$xKN$ of manganese are in an ionized state (${\rm Mn^-_{Ga}}$),
while $x(1-K)N$ is the number of neutral magnetic acceptors ${\rm
Mn^0_{Ga}}$.

It is well known, that the half-filled $d$ shell of manganese acts
as effective spin $S=5/2$. However, our above-mentioned
results~\cite{Kokurin2024} on FM and AFM indirect exchange alignment
of Mn $d$-shell spins in Mn-Mn pairs were obtained using the
simplified $S=1/2$ model. The same types of indirect exchange
interactions will manifest themselves for ${\rm Mn^0_{Ga}}$--${\rm
Mn^-_{Ga}}$ and ${\rm Mn^0_{Ga}}$--${\rm Mn^0_{Ga}}$ pairs, if we
take into account the real $d$-shell spin of 5/2 and hole angular
momentum of 3/2. This can be understood from simple classical
considerations. For example, the ground state of two holes localized
on two centers (two interacting acceptors) corresponds to a zero
projection of the total angular momentum of holes onto the
intercenter axis~\cite{Averkiev1994,Durst2017}. In this case, the
intracenter exchange of each hole with its own $d$ shell leads to
AFM alignment of $d$-shell spins regardless of the sign of the
intracenter exchange.

From the point of view of magnetic ordering, taking into account the
real spin $S=5/2$ will significantly complicate the calculation,
while apparently not leading to qualitatively new results. In what
follows we will use a simplified treatment, assuming that the spin
of the $d$ shell is equal to 1/2. We believe that all qualitative
results concerning compensation effects will remain unchanged
regardless of the magnitude of the Mn $d$-shell spin or the type of
pair exchange interaction (Heisenberg or Ising) used in the
calculation.

To find the free energy $F=E-TS$, it is necessary to take into
account all possible configurations at given concentration $x$ of
magnetic component (alloying disorder) and given magnetization
$\xi=\frac{N_\uparrow-N_\downarrow}{xN}$ \cite{Note1} (magnetic
disorder). Here $N_\uparrow$ ($N_\downarrow$) is the number of spins
with $\uparrow$ ($\downarrow$) orientation,
$xN=N_\uparrow+N_\downarrow$. One should find the total exchange
energy $E$ taking into account two mechanisms of indirect pair
interaction~\cite{Kokurin2024} for pairs of acceptors being both in
neutral (${\rm Mn^0_{Ga}}$--${\rm Mn^0_{Ga}}$) and neutral and
ionized (${\rm Mn^0_{Ga}}$--${\rm Mn^-_{Ga}}$) states \cite{Note2}.

Let us call ``$j$-distant pairs'' the pairs of neutral-neutral or
ionized-neutral Mn's which have intercenter distance coinciding with
the radius $R_j$ of the $j$th CS. The corresponding coordination
numbers $Z_j$ for the case of the fcc lattice are well known (see,
for instance, Ref.~\cite{Thomas1964}) and can be easily calculated.
The total number of $j$-distant magnetic pairs is equal to
$\frac12Z_jx^2N$.

Now we write down the number of neutral-neutral $N^{00}_{jss'}$ and
ionized-neutral $N^{-0}_{jss'}$ $j$-distant pairs with spin
configurations $ss'$, where $s,s'=\uparrow,\downarrow$. Now, for
clarity, let us consider the number $N^{-0}_{j\uparrow\uparrow}$. It
is proportional to the number of $\uparrow$ spins of ionized
manganese ${\rm Mn_{Ga}^-}$, $KN_\uparrow$, and the average number
of $\uparrow$ spins of neutral manganese ${\rm Mn_{Ga}^0}$ in the
$j$th CS relative to each ionized one, $xZ_j(1-K)(N_\uparrow/xN)$.
The number of corresponding pairs as a function of $\xi$ is given by
\[
N^{-0}_{j\uparrow\uparrow}=\frac14Z_jx^2K(1-K)N(1+\xi)^2.
\]

Other numbers of pairs are found by analogy:
\[
N^{-0}_{j\downarrow\downarrow}=\frac14Z_jx^2K(1-K)N(1-\xi)^2,
\]
\[
N^{-0}_{j\uparrow\downarrow}+N^{-0}_{j\downarrow\uparrow}=\frac12Z_jx^2K(1-K)N(1-\xi^2).
\]

The corresponding contribution into the Ising-type exchange energy
\[
E^\mathrm{FM}=-\frac12\sum_jJ^\mathrm{FM}(R_j)(N^{-0}_{j\uparrow\uparrow}+
N^{-0}_{j\downarrow\downarrow}-N^{-0}_{j\uparrow\downarrow}-N^{-0}_{j\downarrow\uparrow})
\]
is finally given by
\begin{equation}
E^\mathrm{FM}=-\frac12x^2K(1-K)N\xi^2\sum_jZ_jJ^\mathrm{FM}(R_j),
\end{equation}
where $J^\mathrm{FM}(r)$ is the distance-dependent exchange
parameter for FM interaction of the ionized-neutral
pair~\cite{Kokurin2024}.

By analogy, the number of $j$-distant neutral-neutral pairs
$N^{00}_{jss'}$ can be found~\cite{Note3}. The corresponding
contribution into exchange energy reads

\begin{equation}
E^\mathrm{AFM}=\frac14x^2(1-K)^2N\xi^2\sum_jZ_j|J^\mathrm{AFM}(R_j)|,
\end{equation}
where $J^\mathrm{AFM}(r)$ is the exchange parameter for AFM
interaction of a neutral-neutral pair, $J^\mathrm{AFM}(r)<0$. It
should be noted that the correlation in the positions of magnetic
centers is not taken into account in our model. This corresponds to
the so-called Bragg-Williams approximation in the theory of
disordered media.

The total exchange energy, $E=E^\mathrm{FM}+E^\mathrm{AFM}$, finally
can be written in the form
\begin{equation}
E=-\frac12xNJ\xi^2,
\end{equation}
with
\[
J=\frac12x(1-K)\sum_jZ_j\left[2KJ^\mathrm{FM}(R_j)-(1-K)|J^\mathrm{AFM}(R_j)|\right].
\]
It should be noted that all parameters of interest (Mn content, CD,
and the lattice structure through $R_j$ and $Z_j$, as well as the
pair exchange interactions) are included in a single parameter $J$.

Let us now calculate the entropy $S$ of the system. The statistical
weight, i.e., the number of configurations corresponding to fixed
magnetization $\xi$, is given by \cite{Note4}

\begin{equation}
P^m(\xi)=\frac{(xN)!}{N_\uparrow!N_\downarrow!}=\frac{(xN)!}{\left(xN\frac{1+\xi}{2}\right)!\left(xN\frac{1-\xi}{2}\right)!}.
\end{equation}

The magnetization-dependent part of free energy has the form
\cite{Note5}

\begin{equation}
\label{free_energy}
F=-\frac12xNJ\xi^2+\frac12xNk_BT[(1+\xi)\ln(1+\xi)+(1-\xi)\ln(1-\xi)].
\end{equation}

The next step is the minimization of free energy $F$ relative to
magnetization ($\partial F/\partial\xi=0$,
$\partial^2F/\partial\xi^2>0$). This leads to the following
equation:
\begin{equation}
\label{molecular_field} \tanh\left(\frac{J\xi}{k_BT}\right)=\xi,
\end{equation}
which looks like the usual transcendental equation of the molecular
field theory \cite{Smart1966}. The numerical solution of
Eq.~(\ref{molecular_field}) is well known: the remanent
magnetization is maximal ($\xi=1$) at zero temperature and
monotonically decreases to zero when the temperature tends to $T_C$.
The Curie temperature is easily determined:
\begin{eqnarray}
\label{T_C} \nonumber k_BT_C&=&\frac12x(1-K)
\\
&\times&\sum_jZ_j\left[2KJ^\mathrm{FM}(R_j)-(1-K)|J^\mathrm{AFM}(R_j)|\right].
\end{eqnarray}

Obviously, Eq.~(\ref{T_C}) makes sense only if the right-hand side
is positive. One can see that at $K\rightarrow 1$ the Curie
temperature $T_C$ tends to zero, which corresponds to the absence of
carriers mediating exchange interaction (overcompensation)
\cite{Kokurin2024}. A transition to the paramagnetic (PM) phase
occurs. At low values of $K$, the Curie temperature is formally
negative. This corresponds to the predominance of AFM pair
interactions. However, the AFM ordering does not appear due to a
random alloy composition. Thus the low-compensated samples can be
considered to be in the spin-glass-like (SG-like) phase.

\section{Discussion}
\label{Discussion}

Let us discuss the dependence $T_C(K)$ given by Eq.~(\ref{T_C}). The
dependence $T_C(K)$ is described by inverted parabola. This means
that there is an intermediate CD value at which $T_C$ reaches
maximum:
\begin{equation}
\label{K_max}
K_{max}=\frac{\sum_jZ_j[J^\mathrm{FM}(R_j)+|J^\mathrm{AFM}(R_j)|]}{\sum_jZ_j[2J^\mathrm{FM}(R_j)+|J^\mathrm{AFM}(R_j)|]}.
\end{equation}

The solution of equation $T_C(K)=0$ leads to the threshold CD value
$K_{th}$:
\begin{equation}
\label{K_th}
K_{th}=\frac{\sum_jZ_j|J^\mathrm{AFM}(R_j)|}{\sum_jZ_j[2J^\mathrm{FM}(R_j)+|J^\mathrm{AFM}(R_j)|]},
\end{equation}
which corresponds to the onset of magnetic ordering.

The further analysis of Eq.~(\ref{T_C}) requires the use of specific
dependencies of exchange parameters $J^\mathrm{FM}(r)$ and
$J^\mathrm{AFM}(r)$. Here the model dependencies from
Ref.~\cite{Kokurin2024} are used (see Appendix D of
Ref.~\cite{Kokurin2024}).

It should be noted that the series in Eq.~(\ref{T_C}) are
convergent, which is due to exponential decay of $J^\mathrm{FM}(r)$
and $J^\mathrm{AFM}(r)$ dependencies. However, the results of
summing of the infinite series leads to very high magnitude of
$T_C$. There are two physical circumstances, which require taking
into account the finite number of terms in the sums of
Eq.~(\ref{T_C}). First, there is the intercenter distance $R_T$ for
which $k_BT\sim|J^i(R_T)|$ ($i=$FM, AFM). This means that definite
$d$-shell orientation, say a codirectional one for the case of FM
alignment, is no longer predominant at such large distances. Thus
the contributions to exchange energy from pairs with $R_j>R_T$
should be excluded. Second, dependencies used~\cite{Kokurin2024}
were obtained in the limit when only one interacting pair in the
crystal is considered and there are no other centers which can
renormalize exchange interaction in the spirit of screening. The
similar ``screening'' is important at distances exceeding an average
intercenter distance corresponding to the Mn content $x$. The second
topic seems more significant.

The simplest way out is to cut off the summation at some threshold
CS ($j=j_{max}$). In this case, the maximal $j$ corresponds to CS
radius $R_j$, which coincides with the characteristic ``screening''
length [cf. corresponding $j_{max}$ found from comparison with
experimental data by means of Eq.~(\ref{fit}) in the next section].
This characteristic length is obviously $x$ dependent. Thus the
simple cutoff procedure will lead to discontinuities of $T_C$ as a
function of $x$.

There is another, more consistent approach. For example, the
following continuous distribution function~\cite{Williams1968}
\begin{equation}
\label{distrib} G(R)=4\pi R^2N\exp\left(-\frac43\pi R^3N\right)
\end{equation}
is used to account for the random distribution of the donor-acceptor
pair, where $N$ is now the concentration of ions. Such a
consideration accounts for only nearest-neighbor pairs, i.e., $G(R)$
gives the probability of the realization of $R$-distant pairs, when
there are no other centers inside a sphere of radius $R$. A discrete
analog of the above distribution is used here.

When deriving Eq.~(\ref{T_C}), we used the average numbers of the
$j$-distant pairs $N_j=\sum_{n=0}^{Z_j}P_{Z_j}(n)n=xZ_j$ for each
magnetic center, which are found with the help of the binomial
distribution (BD)
\begin{equation}
P_{Z_j}(n)=\frac{Z_j!}{n!(Z_j-n)!}x^n(1-x)^{Z_j-n},
\end{equation}
which determines the probability to occupy $n$ sites of $j$th CS
with magnetic ions. One can modify such a probability function
accounting for the $j$-distant pairs at simultaneous presence at
inner CSs [from first to $(j-1)$th] of no more than $M$ magnetic
centers. In this case, the characteristic ``screening'' length is of
the order of the CS radius, within which there are, on average, $M$
magnetic centers. The numbers $N_j$ differ from $xZ_j$ by a factor
\begin{equation}
P^{(j-1)}(M)=\sum_{l=0}^{M}P_{Z^{(j-1)}}(l),
\end{equation}
where $Z^{(j-1)}=\sum_{k=1}^{j-1}Z_k$ is the number of sites inside
the $j$th CS. Such a function is close to 1 for the first few CSs,
but rapidly decreases when $j$ increases. Thus the sum in
Eq.~(\ref{T_C}) should be calculated by replacing $xZ_j$ with $N_j$.
The special case $M=0$ corresponds to nearest-neighbor
approximation, and it is a discrete analog of distribution
(\ref{distrib}). At the same time, it is obvious that the
nearest-neighbor approximation is not enough in a diluted system and
Curie temperature found in this approximation is underestimated.

\begin{figure*}
\includegraphics[width=\linewidth]{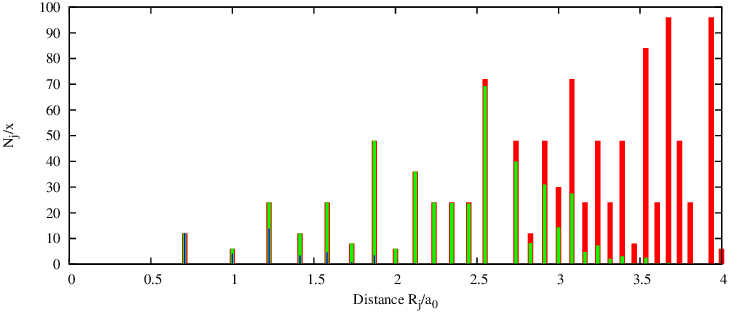}
\caption{\label{fig01} Average number of $j$-distant pairs $N_j$ vs
CS radius $R_j$ in units of the lattice constant $a_0$ (Mn content
$x=0.03$). Red lines correspond to the ``nonscreened'' case,
$N_j=xZ_j$. Green lines account for a partial ``screening'' with
$M=Z_1=12$. Blue lines correspond to complete ``screening'' ($M=0$,
nearest-neighbor approximation).}
\end{figure*}

The number of $j$-distant pairs taken into account when calculating
$T_C$ is depicted in Fig.~\ref{fig01} for different ``screening''
levels. The fcc structure of the cation sublattice is taken into
account. One can see that the above-mentioned ``screening'' is
nonessential for inner CSs and it is important for CSs, for which
$Z^{(j-1)}$ satisfies an inequality $xZ^{(j-1)}\gtrsim M$.

\begin{figure}
\includegraphics{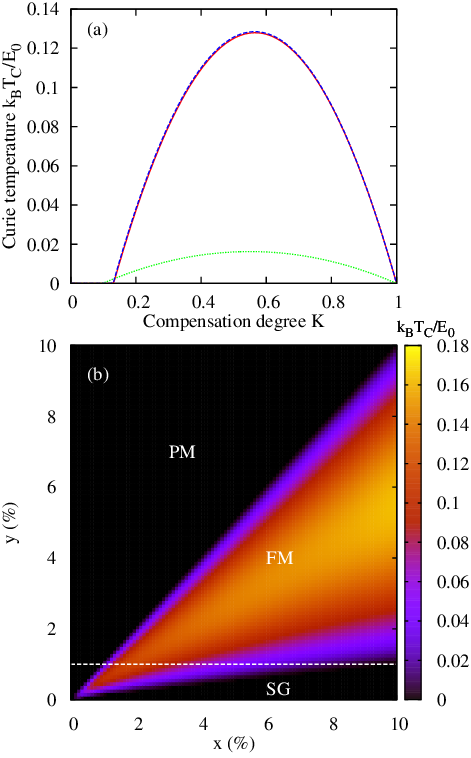}
\caption{\label{fig02} (a) Dependence of Curie temperature $T_C$ vs
CD $K$ at $x=0.03$. The different lines correspond to different
accounting of ``screening'': cutoff with $j_{max}=17$ (red solid
line), ``screening'' with $M=12$ (blue dashed line), and
nearest-neighbor approximation $M=0$ (green dotted line). (b) The
Curie temperature ($k_BT_C/E_0$) as a function of the Mn content $x$
and the content of compensating donors $y$. The areas corresponding
to different magnetic phases (PM, FM, and SG-like) are marked. The
horizontal dashed line indicates the position of the abscissa axis
when compensation by As antisites is taken into account. The
intracenter exchange constant is $A=-0.2E_0$.}
\end{figure}

The dependence of Curie temperature $T_C$ on CD $K$ at fixed
magnetic content $x$ is depicted in Fig.~\ref{fig02}(a). Different
types of ``screening'' are used. The maximum $T_C$ in the case of
nearest-neighbor ($M=0$) and other two types of ``screening''
($j_{max}=17$ and $M=12$) differs almost in order of magnitude
($\sim 7$ times). It should be noted that the summation of infinite
series of Eq.~(\ref{T_C}) leads to an even greater value of $T_C$
(more than an order of magnitude compared to the case $M=12$). The
$T_C(K)$ dependence is of threshold type, i.e., nonzero $T_C$ (FM
ordering) arises at finite CD $K_{th}$, which is itself a function
of $x$. The maximum of $T_C$ corresponds to definite $K$ ($1/2\leq
K_{max}\leq1$).

It should be noted that the numerical calculation using model
dependencies~\cite{Kokurin2024} of pair exchange parameters does not
claim to provide the exact value of $T_C$, but it gives qualitative
behavior of $T_C(x,K)$ dependence. However, the numerical estimate
gives for $T_C$ the value about 140 K at maximum [$k_BT_C/E_0\simeq
0.128$ in Fig.~\ref{fig02}(a)] with the use of $E_0=96.9$ meV
\cite{Note6} Here, $E_0$ is the ${\rm Mn_{Ga}^0}$ acceptor binding
energy without taking into account the exchange contribution.

Consideration within the IB model means that the indirect exchange
parameters are related to the overlap of the wave functions of holes
localized at different centers. In this case, one should expect the
usual exponential decrease of the exchange parameters $J^i(r)$
($i$=FM, AFM) without any alternating features. It may seem that,
when taking into account the influence of the surrounding Mn
centers, the sign of $J^i(r)$ can change depending on the
configuration of Mn impurities around a given pair. At the same
time, intuitively, self-averaging will lead to a simple
renormalization in the spirit of ordinary dielectric screening, when
the permittivity $\varepsilon$ appears. It should be noted here that
in the nearest-neighbor approximation ($M=0$) the picture of
pairwise interactions is accurate. A gradual increase in the
parameter $M$ simply leads to renormalization and no features arise
in the $T_C(K)$ dependence.

When deriving the expression for the exchange energy, we assumed
that the pair exchange parameters $J^i(r)$ ($i=$FM, AFM) are
functions of distance only. The dependence on $x$ and $K$ appears in
the effective exchange constant $J$ through the weighting factors at
$J^i(r)$. This approximation corresponds to a small IB width. The
corresponding Mn concentration can be easily estimated. The Bohr
radius of the ${\rm Mn_{Ga}^0}$ center is about 1 nm. The Mn
concentration corresponding to such an average distance between
acceptors is $10^{21}$ cm$^{-3}$. This corresponds to a manganese
content of about 5\%. The finite IB width can be taken into account,
which will lead to additional spread not only in the binding energy
of ${\rm Mn_{Ga}^0}$ acceptors, but also in the $J^i(r)$
dependencies. This will require additional averaging, the effect of
which will most likely be little different from the ``screening''
effect.

The dependence of Curie temperature on Mn content $x$ and on the
content of compensating nonmagnetic donors $y$ is depicted in
Fig.~\ref{fig02}(b). In Fig.~\ref{fig02}(b) the line $y(x)$,
corresponding to maximal $T_C(x)$ values, has a sublinear
dependence. The dependence $T_C(x)$ along this line is sublinear as
well: the $T_C$ value corresponding to $x=0.1$ at $K=K_{max}$ is
about 200 K (cf. with 140 K at $x=0.03$).

The map of Fig.~\ref{fig02}(b) looks like a phase diagram. The area
$y>x$ ($K>1$, $K=y/x$) corresponds to the PM phase. The area with
positive $T_C$ corresponds to FM ordering. The SG-like phase is
realized if the AFM exchange predominates (formally $T_C<0$).

At first glance there is a FM phase even at very low $x$. However,
the additional compensation due to ${\rm As_{Ga}}$ antisites (double
donors), which are inevitable at low-temperature growth, is not
taken into account in our consideration. The account of additional
compensation through the antisite content $x_{As}$ provides a simple
shift of the Fig.~\ref{fig02}(b) map down along the $y$ axis by the
$2x_{As}$ value. In this case, the threshold Mn content
$x_{th}=2x_{As}$ arises even at $y=0$, i.e., the minimal Mn content
$2x_{As}$ is necessary for the emergence of the FM phase. The
antisite concentration can exceed $10^{20}$
cm$^{-3}$~\cite{Liu1995,Grandidier2000}, which corresponds to
$x_{As}\sim 0.5\%$. The latter is in agreement with minimum Mn
content of $1\%$ required for the onset of ferromagnetism in
(Ga,Mn)As~\cite{Muneta2013}. Thus, within our approach,
Ga$_{1-x}$Mn$_x$As is in a PM state both at low and at very high Mn
content. This occurs due to overcompensation by As antisites in the
first case and by Mn interstitials in the second. Taking into
account the additional compensation due to the As antisites also
significantly reduces the region corresponding to the SG-like phase
[see Fig.~\ref{fig02}(b)]. Moreover, $x$ above 6\% corresponds to a
concentration of Mn interstitials $x_I$ exceeding
1.5\%~\cite{Jungwirth2005}. This would correspond to compensation
due to nonmagnetic donors $y$ in Fig.~\ref{fig02}(b) above 3\%. Thus
the SG-like phase in (Ga,Mn)As is apparently not realized.

It should be noted that the obtained distribution (see
Fig.~\ref{fig01}), which is a discrete generalization of
distribution~(\ref{distrib}), is of a phenomenological nature and
its microscopic justification is questionable. A similar
distribution can be used in other problems related to disordered
solids, where pairwise interactions take place. Taking into account
the various screening parameters $M$ is close to a simple
renormalization of the curves in Fig.~\ref{fig02}(a). At the same
time, there are a number of quantities that do not depend (or depend
weakly) on $M$. These are the CD values corresponding to the maximum
value of $T_C$ and the ferromagnetism threshold, $K_{max}$ and
$K_{th}$, respectively [see Eqs.~(\ref{K_max}) and (\ref{K_th}) with
the replacement $xZ_j\rightarrow N_j$].

Our numerical results do not claim to provide a quantitative
description of magnetic ordering. They have a qualitative or
semiquantitative character, since used dependencies
$J^\mathrm{FM}(r)$ and $J^\mathrm{AFM}(r)$ are derived from rather
simplified model calculations \cite{Kokurin2024}. The more realistic
dependencies $J^\mathrm{FM}(r)$ and $J^\mathrm{AFM}(r)$ can be
derived from consideration of the spin structure of two-center Mn
complexes with one (${\rm Mn}_2^-$) and two (${\rm Mn}_2^0$) holes,
when a realistic VB structure ($\Gamma_8$ symmetry) is taken into
account. The energy and spin structure of such complexes can be
determined within our approach \cite{Kokurin2024} applied to
nonmagnetic acceptor complexes with one (${\rm A}_2^-$) and two
(${\rm A}_2^0$) holes (see Refs.~\cite{Kokurin2023} and
\cite{Durst2017}, respectively).

Let us discuss the photoferromagnetic (PFM) effect
\cite{Karpenko1964,Berdyshev1966} as it applies to (Ga,Mn)As. We
predict a new mechanism of the PFM effect based on the possibility
to control an effective CD by optical means. The
photoionization~\cite{Chapman1967,Brown1973} or
photoneutralization~\cite{Kokurin2013,Petrov2016} excitation scheme
can be used. Increasing the intensity of optical excitation leads to
different behavior of the effective CD (decreasing or increasing) in
the case of photoneutralization or photoionization. The dependence
of Curie temperature on the excitation intensity depends not only on
the excitation mechanism but on the dark CD magnitude relative to
$K_{max}$. It should be noted that the experimental variation of
Curie temperature at the optical excitation of photoneutralization
(photoionization) transition will serve as direct evidence of the
described mechanism of ferromagnetism.

\section{Comparison with experimental data}
\label{Compare}

Let us discuss some experimental results from the point of view of
proposed theory. The $p$-type doping of layer, grown after the
magnetic one, with nonmagnetic acceptor (Be) leads to specific $T_C$
behavior~\cite{Cho2008}. At small Be content $N_{\rm Be}$ the Curie
temperature increases with it. However, as $N_{\rm Be}$ increases
further, this trend reverses: at some point, $T_C$ begins to
decrease, and eventually the ferromagnetism of (Ga,Mn)As disappears
completely. In terms of our theory this means that initially the
Be-undoped sample is highly compensated ($K>K_{max}$). The $p$-type
doping leads to the CD sequential decreasing in magnetic layer. The
Curie temperature increases, reaches its maximum value at
$K=K_{max}$, and decreases at $K<K_{max}$ [see Fig.~\ref{fig02}(a)].

The results of Ref.~\cite{Cho2008a} concerning the dependence of
$T_C$ on Si counterdoping can be explained in terms of our theory as
well. The Si-doped samples exhibit the increase of remanent
magnetization at $T\rightarrow 0$ compared to Si-undoped
Ga$_{1-x}$Mn$_x$As samples both at low and high Mn content. This
corresponds to decreasing of ${\rm Mn_I}$ content and, in turn,
weakening of magnetic compensation ($x_{eff}$ increases). At the
same time, the effect of counterdoping on Curie temperature is
different for the samples with low and high Mn content $x$. At low
$x$ the Curie temperature decreases compared to the Si-undoped
samples with the same $x$, whereas at high $x$ it increases. This
means that at low (high) $x$ the undoped samples have deficient
(excessive) ${\rm Mn_I}$-mediated charge compensation, i.e.,
$K<K_{max}$ ($K>K_{max}$). As a result, the CD reduction through the
Si doping leads to decreasing (increasing) of Curie temperature [see
Fig.~\ref{fig02}(a)] compared to undoped samples.

\begin{table*}
\caption{\label{Table1} Experimental data of
Ref.~\cite{Dobrowolska2012} on content of different Mn defects and
Curie temperature. The Mn content is in units of \%. The data of
Ref.~\cite{Dobrowolska2012} are recalculated into compensation
degree and $T_C/x_{eff}$ as well as corresponding errors (see
Appenidix \ref{AppA} for details). The used errors are $\Delta
x=0.05\%$ and $\Delta T_C=5$ K. The content of As antisites is
$0.45$ \%, which corresponds to a concentration of $1\times 10^{20}$
cm$^{-3}$~\cite{Liu1995,Grandidier2000}.}
\begin{tabular}{ c  c  c  c  c  c  c  c  c  c  c  c  c }
\hline \hline Sample & $x_{tot}$ & $x_{random}$ & $x_{sub}$ & $x_I$
& $x_{eff}$ & $x_{As}$ & $K=(x_I+2x_{As})/x_{eff}$ &
$K^0=x_I/x_{eff}$ & $\Delta K$ & $T_C$ (K) & $T_C/x_{eff}$ (K) &
$\Delta(T_C/x_{eff})$ (K)
 \\ \hline
A & 3.0 & 0.1 & 2.4 & 0.5 & 1.9 & 0.45 & 0.737 & 0.263 &
0.062 & 45 & 2370 & 270  \\
B & 3.5 & 0.1 & 3.2 & 0.2 & 3.0 & 0.45 & 0.367 &
0.067 & 0.038 & 27 & 900 & 167 \\
C & 4.8 & 0.3 & 4.0 & 0.5 & 3.5 & 0.45 & 0.400 &
 0.143 & 0.032 & 34 & 971 & 144 \\
C*\footnote{Annealed samples are marked with an asterisk.} & 5.0 &
0.4 & 4.3 & 0.3 & 4.0 & 0.45 & 0.300 &
 0.075 & 0.028 & 54 & 1350 & 126\\
D & 5.8 & 0.3 & 5.0 & 0.5 & 4.5 & 0.45 & 0.311 & 0.111 &
0.025 & 26 & 578 & 111 \\
D* & 5.9 & 0.4 & 5.3 & 0.2 & 5.1 & 0.45 & 0.216 & 0.039 &
0.022 & 42 & 824 & 98\\
E & 6.0 & 0.3 & 5.0 & 0.7 & 4.3 & 0.45 & 0.372 & 0.163 &
0.026 & 30 & 698 & 117\\
E* & 6.1 & 0.5 & 5.2 & 0.4 & 4.8 & 0.45 & 0.271 & 0.083 &
0.023 & 43 & 896 & 105\\
F & 6.4 & 0.5 & 4.2 & 1.7 & 2.5 & 0.22\footnote{This $x_{As}$ value
is halved because at $x_{As}=0.45$\% the value of $K$ exceeds 1. The
latter corresponds to the absence of the magnetic ordering within
our approach. The above reduction is justified since an increase in
the ${\rm Mn_{Ga}}$ content leads to a decrease in the concentration
of As antisites.} & 0.856 & 0.680 &
0.048 & 40 & 1600 & 203\\
F* & 6.8 & 1.3 & 4.5 & 1.0 & 3.5 & 0.45 & 0.543 & 0.286 &
0.033 & 90 & 2571 & 148\\
\hline \hline
  \end{tabular}
\end{table*}

The above description of the magnetic properties of additionally
doped (Ga,Mn)As, based on our theory, is rather qualitative in
nature. This is due to the fact that most experimental data contain
information about total Mn content only, rather than that of
different Mn charge states and other defects. This means that the
real CD cannot be estimated with any degree of accuracy and, in
turn, the quantitative comparison of theory and experiment is
impossible. To our knowledge, the data on the content of various Mn
charge states are presented only in Ref.~\cite{Dobrowolska2012}.
This allows one to estimate CD (see Table~\ref{Table1}) and make a
direct comparison with the present theory. The additional source of
electrical compensation through ${\rm As_{Ga}}$
antisites~\cite{Liu1995,Grandidier2000} is taken into account.

To fit the experimental dependence of $T_C/x_{eff}$ on $K$,
Eq.~(\ref{T_C}) is rewritten in the form:
\begin{equation}
\label{fit}
\frac{k_BT_C}{x_{eff}}=(1-K)\bigg[Ka-\frac12(1-K)b\bigg],
\end{equation}
with
\[
a=\sum_jZ_jJ^\mathrm{FM}(R_j),\qquad
b=\sum_jZ_j|J^\mathrm{AFM}(R_j)|,
\]
acting now as fitting parameters.

\begin{figure}[t]
\includegraphics[width=\linewidth]{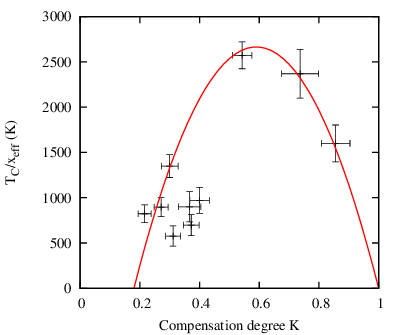}
\caption{\label{fig03} Experimental data of
Ref.~\cite{Dobrowolska2012} as well as the calculated theoretical
curve for the dependence of $T_C/x_{eff}$ on CD $K$. The theoretical
curve corresponds to the fit by means of Eq.~(\ref{fit}) with
$a/k_B=13000$ K and $b/k_B=5700$ K ($a=1.12$ eV and $b=0.49$ eV).}
\end{figure}

Equation~(\ref{fit}), containing fitting parameters (determined by
the strength and action range of two competing exchange mechanisms),
although obtained in the approximation $S=1/2$, in our opinion is
quite suitable for comparison with experimental data. The
experimental points from Ref.~\cite{Dobrowolska2012} (see
Table~\ref{Table1}) together with the theoretical fit (\ref{fit})
are depicted in Fig.~\ref{fig03}. A similar trend is observed in
both experimental and theoretical data. The fitting parameters ($a$
and $b$) allow one to estimate the ``screening'' length (the maximum
distance between magnetic ions contributing into ferromagnetism) as
close to $3a_0$~\cite{Note7}. At the same time, Fig.~\ref{fig03}
does not claim to be an exact comparison of theory and experiment.
The main obstacle is the absence of exact data on the ${\rm
As_{Ga}}$ antisites content. The latter can significantly change the
compensation degree (cf. the data of $K$ and $K^0$ columns in
Table~\ref{Table1}) and, in turn, the position of experimental
points. For example, the halving of $x_{As}$ for experimental points
lying under the parabola (samples B, C, D, and E) leads to a
coincidence with the theoretical curve. When evaluating the
experimental values of CD (Table~\ref{Table1}), the antisite content
is taken to be $x$ independent. However, it seems the antisite
content $x_{\rm As}$ is $x$ dependent in reality. It should be noted
that, for a real comparison of theory and experiment, detailed
experimental data are required: in particular, more precise
information on the content of various defects affecting the
compensation level.

\section{Conclusion and outlook}
\label{Conclusion}

In conclusion, the consistent IB-based theory of ferromagnetism in
the (Ga,Mn)As alloy is presented. The compensation is a key
parameter of the problem along with the content of the magnetic
component. The molecular-field-like equation, which describes the
temperature dependence of remanent magnetization, is derived. All
the necessary information about the material is hidden in a single
parameter. For a fixed Mn content, the Curie temperature can vary
significantly with CD. Moreover, the magnetic phase transition can
occur when CD changes. The theory, which considers the concentration
of the compensating impurity $y$ as an independent parameter,
predicts the existence of FM, PM, and SG-like phases. However,
taking into account the compensation due to As antisites and the
appearance of Mn interstitials at high Mn content $x$ indicates the
absence of the SG-like phase in real (Ga,Mn)As alloys. The predicted
PFM effect may serve as a confirmation of the proposed mechanism of
magnetic ordering in (Ga,Mn)As. It should be noted that the CD
variation and, in turn, the magnetic state control in DMS are
possible by means of electric gates as well.

The presented theory allows one to explain the behavior of remanent
magnetization and Curie temperature in additionally $p$- and
$n$-doped (Ga,Mn)As samples through the CD variation. The
predictions of our theory were compared with the results of
Ref.~\cite{Dobrowolska2012}, where the separate concentrations of
different Mn charge states are presented. Other experimental works
contain, as a rule, the information about total Mn content $x$ only,
which is not enough for comparison. This problem can be solved in
two different ways: (i) the detail resolution of different Mn states
and other compensating defects with determination of corresponding
concentrations, which will make it possible to determine the CD
value; (ii) the development of advanced theory, containing
additional thermodynamical calculations, which allow one to find
separate concentrations for different defects at fixed $x$. Progress
in both directions is likely to yield fruitful results.

Our model, as well as Zener-type models, formally predict an
increase in the Curie temperature with increasing $x$. Experimental
data point out the absence of such an opportunity. This means that
the above models do not take into account some important
circumstances. The probability that a substituting Mn will have a
nearest-neighbor Mn in the cation sublattice is $p=1-(1-x)^{Z_1}$
(the first CS has $R_1=a_0/\sqrt{2}$ and $Z_1=12$ in the case of the
fcc lattice). At $x=0.05$ the probability is $p=0.46$, whereas at
$x=0.1$ it is already $0.72$. This means that at such concentrations
the nearest-neighbor superexchange, which is of AFM type, already
plays a significant role. Thus it is highly likely the AFM
superexchange is the main obstacle for $T_C$ increasing, rather than
compensation, which we have shown to be necessary to some extent.

\begin{acknowledgements}
We acknowledge financial support from the Russian Science Foundation
(Grant No. 25-12-00093).
\end{acknowledgements}

\section*{Data availability}

There are no publicly available research data or software supporting
this manuscript. Requests for further information or data should be
sent to the authors.

\appendix*

\section{Details of experimental data processing}
\label{AppA}

In the case of indirect measurements of quantity $y$, which is a
function $y=f(x_1,x_2,...,x_N)$ of measured values $x_i$
($i=1,2,...,N$), the corresponding error is given by

\begin{equation}
\Delta y=\sqrt{\sum_i\left(\frac{\partial f}{\partial x_i}\Delta
x_i\right)^2}.
\end{equation}

This standard approach leads to the following error of
$T_C/x_{eff}$:
\begin{equation}
\Delta\left(\frac{T_C}{x_{eff}}\right)=\left(\frac{T_C}{x_{eff}}\right)\sqrt{\left(\frac{\Delta
T_C}{T_C}\right)^2+\left(\frac{\Delta x}{x_{eff}}\right)^2}.
\end{equation}

The CD value depends on the content of ${\rm Mn_I}$, ${\rm
Mn_{Ga}}$, and ${\rm As_{Ga}}$: $K=(x_I+2x_{As})/x_{eff}$. The
corresponding error is given by
\begin{equation}
\Delta K=\left(\frac{\Delta x}{x_{eff}}\right)\sqrt{(K^2+5)}.
\end{equation}

Here, the equal values for $\Delta x_i$ (i=$I$, ${\rm As}$, and
eff) are used, $\Delta x_i=\Delta x$.


\begin{thebibliography}{46}%
\makeatletter
\providecommand \@ifxundefined [1]{%
 \@ifx{#1\undefined}
}%
\providecommand \@ifnum [1]{%
 \ifnum #1\expandafter \@firstoftwo
 \else \expandafter \@secondoftwo
 \fi
}%
\providecommand \@ifx [1]{%
 \ifx #1\expandafter \@firstoftwo
 \else \expandafter \@secondoftwo
 \fi
}%
\providecommand \natexlab [1]{#1}%
\providecommand \enquote  [1]{``#1''}%
\providecommand \bibnamefont  [1]{#1}%
\providecommand \bibfnamefont [1]{#1}%
\providecommand \citenamefont [1]{#1}%
\providecommand \href@noop [0]{\@secondoftwo}%
\providecommand \href [0]{\begingroup \@sanitize@url \@href}%
\providecommand \@href[1]{\@@startlink{#1}\@@href}%
\providecommand \@@href[1]{\endgroup#1\@@endlink}%
\providecommand \@sanitize@url [0]{\catcode `\\12\catcode
`\$12\catcode
  `\&12\catcode `\#12\catcode `\^12\catcode `\_12\catcode `\%12\relax}%
\providecommand \@@startlink[1]{}%
\providecommand \@@endlink[0]{}%
\providecommand \url  [0]{\begingroup\@sanitize@url \@url }%
\providecommand \@url [1]{\endgroup\@href {#1}{\urlprefix }}%
\providecommand \urlprefix  [0]{URL }%
\providecommand \Eprint [0]{\href }%
\providecommand \doibase [0]{http://dx.doi.org/}%
\providecommand \selectlanguage [0]{\@gobble}%
\providecommand \bibinfo  [0]{\@secondoftwo}%
\providecommand \bibfield  [0]{\@secondoftwo}%
\providecommand \translation [1]{[#1]}%
\providecommand \BibitemOpen [0]{}%
\providecommand \bibitemStop [0]{}%
\providecommand \bibitemNoStop [0]{.\EOS\space}%
\providecommand \EOS [0]{\spacefactor3000\relax}%
\providecommand \BibitemShut  [1]{\csname bibitem#1\endcsname}%
\let\auto@bib@innerbib\@empty
\bibitem [{\citenamefont {Ohno}\ \emph {et~al.}(1996)\citenamefont {Ohno},
  \citenamefont {Shen}, \citenamefont {Matsukura}, \citenamefont {Oiwa},
  \citenamefont {Endo}, \citenamefont {Katsumoto},\ and\ \citenamefont
  {Iye}}]{Ohno1996}%
  \BibitemOpen
  \bibfield  {author} {\bibinfo {author} {\bibfnamefont {H.}~\bibnamefont
  {Ohno}}, \bibinfo {author} {\bibfnamefont {A.}~\bibnamefont {Shen}}, \bibinfo
  {author} {\bibfnamefont {F.}~\bibnamefont {Matsukura}}, \bibinfo {author}
  {\bibfnamefont {A.}~\bibnamefont {Oiwa}}, \bibinfo {author} {\bibfnamefont
  {A.}~\bibnamefont {Endo}}, \bibinfo {author} {\bibfnamefont {S.}~\bibnamefont
  {Katsumoto}}, \ and\ \bibinfo {author} {\bibfnamefont {Y.}~\bibnamefont
  {Iye}},\ }\bibfield  {title} {\enquote {\bibinfo {title} {({G}a,{M}n){A}s:
  {A} new diluted magnetic semiconductor based on {G}a{A}s},}\ }\href {\doibase
  10.1063/1.118061} {\bibfield  {journal} {\bibinfo  {journal} {Appl. Phys.
  Lett.}\ }\textbf {\bibinfo {volume} {69}},\ \bibinfo {pages} {363} (\bibinfo
  {year} {1996})}\BibitemShut {NoStop}%
\bibitem [{\citenamefont {Dietl}\ \emph {et~al.}(2000)\citenamefont {Dietl},
  \citenamefont {Ohno}, \citenamefont {Matsukura}, \citenamefont {Cibert},\
  and\ \citenamefont {Ferrand}}]{Dietl2000}%
  \BibitemOpen
  \bibfield  {author} {\bibinfo {author} {\bibfnamefont {T.}~\bibnamefont
  {Dietl}}, \bibinfo {author} {\bibfnamefont {H.}~\bibnamefont {Ohno}},
  \bibinfo {author} {\bibfnamefont {F.}~\bibnamefont {Matsukura}}, \bibinfo
  {author} {\bibfnamefont {J.}~\bibnamefont {Cibert}}, \ and\ \bibinfo {author}
  {\bibfnamefont {D.}~\bibnamefont {Ferrand}},\ }\bibfield  {title} {\enquote
  {\bibinfo {title} {Zener model description of ferromagnetism in zinc-blende
  magnetic semiconductors},}\ }\href {\doibase 10.1126/science.287.5455.1019}
  {\bibfield  {journal} {\bibinfo  {journal} {Science}\ }\textbf {\bibinfo
  {volume} {287}},\ \bibinfo {pages} {1019} (\bibinfo {year}
  {2000})}\BibitemShut {NoStop}%
\bibitem [{\citenamefont {Dietl}\ \emph {et~al.}(2001)\citenamefont {Dietl},
  \citenamefont {Ohno},\ and\ \citenamefont {Matsukura}}]{Dietl2001}%
  \BibitemOpen
  \bibfield  {author} {\bibinfo {author} {\bibfnamefont {T.}~\bibnamefont
  {Dietl}}, \bibinfo {author} {\bibfnamefont {H.}~\bibnamefont {Ohno}}, \ and\
  \bibinfo {author} {\bibfnamefont {F.}~\bibnamefont {Matsukura}},\ }\bibfield
  {title} {\enquote {\bibinfo {title} {Hole-mediated ferromagnetism in
  tetrahedrally coordinated semiconductors},}\ }\href {\doibase
  10.1103/PhysRevB.63.195205} {\bibfield  {journal} {\bibinfo  {journal} {Phys.
  Rev. B}\ }\textbf {\bibinfo {volume} {63}},\ \bibinfo {pages} {195205}
  (\bibinfo {year} {2001})}\BibitemShut {NoStop}%
\bibitem [{\citenamefont {Jungwirth}\ \emph {et~al.}(2005)\citenamefont
  {Jungwirth}, \citenamefont {Wang}, \citenamefont {Ma\ifmmode~\check{s}\else
  \v{s}\fi{}ek}, \citenamefont {Edmonds}, \citenamefont {K\"onig},
  \citenamefont {Sinova}, \citenamefont {Polini}, \citenamefont {Goncharuk},
  \citenamefont {MacDonald}, \citenamefont {Sawicki}, \citenamefont
  {Rushforth}, \citenamefont {Campion}, \citenamefont {Zhao}, \citenamefont
  {Foxon},\ and\ \citenamefont {Gallagher}}]{Jungwirth2005}%
  \BibitemOpen
  \bibfield  {author} {\bibinfo {author} {\bibfnamefont {T.}~\bibnamefont
  {Jungwirth}}, \bibinfo {author} {\bibfnamefont {K.~Y.}\ \bibnamefont {Wang}},
  \bibinfo {author} {\bibfnamefont {J.}~\bibnamefont {Ma\ifmmode~\check{s}\else
  \v{s}\fi{}ek}}, \bibinfo {author} {\bibfnamefont {K.~W.}\ \bibnamefont
  {Edmonds}}, \bibinfo {author} {\bibfnamefont {J\"urgen}\ \bibnamefont
  {K\"onig}}, \bibinfo {author} {\bibfnamefont {Jairo}\ \bibnamefont {Sinova}},
  \bibinfo {author} {\bibfnamefont {M.}~\bibnamefont {Polini}}, \bibinfo
  {author} {\bibfnamefont {N.~A.}\ \bibnamefont {Goncharuk}}, \bibinfo {author}
  {\bibfnamefont {A.~H.}\ \bibnamefont {MacDonald}}, \bibinfo {author}
  {\bibfnamefont {M.}~\bibnamefont {Sawicki}}, \bibinfo {author} {\bibfnamefont
  {A.~W.}\ \bibnamefont {Rushforth}}, \bibinfo {author} {\bibfnamefont {R.~P.}\
  \bibnamefont {Campion}}, \bibinfo {author} {\bibfnamefont {L.~X.}\
  \bibnamefont {Zhao}}, \bibinfo {author} {\bibfnamefont {C.~T.}\ \bibnamefont
  {Foxon}}, \ and\ \bibinfo {author} {\bibfnamefont {B.~L.}\ \bibnamefont
  {Gallagher}},\ }\bibfield  {title} {\enquote {\bibinfo {title} {Prospects for
  high temperature ferromagnetism in ({G}a,{M}n){A}s semiconductors},}\ }\href
  {\doibase 10.1103/PhysRevB.72.165204} {\bibfield  {journal} {\bibinfo
  {journal} {Phys. Rev. B}\ }\textbf {\bibinfo {volume} {72}},\ \bibinfo
  {pages} {165204} (\bibinfo {year} {2005})}\BibitemShut {NoStop}%
\bibitem [{\citenamefont {Ohya}\ \emph {et~al.}(2011)\citenamefont {Ohya},
  \citenamefont {Takata},\ and\ \citenamefont {Tanaka}}]{Ohya2011}%
  \BibitemOpen
  \bibfield  {author} {\bibinfo {author} {\bibfnamefont {S.}~\bibnamefont
  {Ohya}}, \bibinfo {author} {\bibfnamefont {K.}~\bibnamefont {Takata}}, \ and\
  \bibinfo {author} {\bibfnamefont {M.}~\bibnamefont {Tanaka}},\ }\bibfield
  {title} {\enquote {\bibinfo {title} {Nearly non-magnetic valence band of the
  ferromagnetic semiconductor {G}a{M}n{A}s},}\ }\href
  {https://www.nature.com/articles/nphys1905} {\bibfield  {journal} {\bibinfo
  {journal} {Nat. Phys.}\ }\textbf {\bibinfo {volume} {7}},\ \bibinfo {pages}
  {342} (\bibinfo {year} {2011})}\BibitemShut {NoStop}%
\bibitem [{\citenamefont {Dobrowolska}\ \emph {et~al.}(2012)\citenamefont
  {Dobrowolska}, \citenamefont {Tivakornsasithorn}, \citenamefont {Liu},
  \citenamefont {Furdyna}, \citenamefont {Berciu}, \citenamefont {Yu},\ and\
  \citenamefont {Walukiewicz}}]{Dobrowolska2012}%
  \BibitemOpen
  \bibfield  {author} {\bibinfo {author} {\bibfnamefont {M.}~\bibnamefont
  {Dobrowolska}}, \bibinfo {author} {\bibfnamefont {K.}~\bibnamefont
  {Tivakornsasithorn}}, \bibinfo {author} {\bibfnamefont {X.}~\bibnamefont
  {Liu}}, \bibinfo {author} {\bibfnamefont {J.~K.}\ \bibnamefont {Furdyna}},
  \bibinfo {author} {\bibfnamefont {M.}~\bibnamefont {Berciu}}, \bibinfo
  {author} {\bibfnamefont {K.~M.}\ \bibnamefont {Yu}}, \ and\ \bibinfo {author}
  {\bibfnamefont {W.}~\bibnamefont {Walukiewicz}},\ }\bibfield  {title}
  {\enquote {\bibinfo {title} {Controlling the {C}urie temperature in
  ({G}a,{M}n){A}s through location of the {F}ermi level within the impurity
  band},}\ }\href {\doibase 10.1038/nmat3250} {\bibfield  {journal} {\bibinfo
  {journal} {Nat. Mater.}\ }\textbf {\bibinfo {volume} {11}},\ \bibinfo {pages}
  {444} (\bibinfo {year} {2012})}\BibitemShut {NoStop}%
\bibitem [{\citenamefont {Muneta}\ \emph {et~al.}(2013)\citenamefont {Muneta},
  \citenamefont {Terada}, \citenamefont {Ohya},\ and\ \citenamefont
  {Tanaka}}]{Muneta2013}%
  \BibitemOpen
  \bibfield  {author} {\bibinfo {author} {\bibfnamefont {I.}~\bibnamefont
  {Muneta}}, \bibinfo {author} {\bibfnamefont {H.}~\bibnamefont {Terada}},
  \bibinfo {author} {\bibfnamefont {S.}~\bibnamefont {Ohya}}, \ and\ \bibinfo
  {author} {\bibfnamefont {M.}~\bibnamefont {Tanaka}},\ }\bibfield  {title}
  {\enquote {\bibinfo {title} {Anomalous {F}ermi level behavior in {G}a{M}n{A}s
  at the onset of ferromagnetism},}\ }\href {\doibase 10.1063/1.4816133}
  {\bibfield  {journal} {\bibinfo  {journal} {Appl. Phys. Lett.}\ }\textbf
  {\bibinfo {volume} {103}},\ \bibinfo {pages} {032411} (\bibinfo {year}
  {2013})}\BibitemShut {NoStop}%
\bibitem [{\citenamefont {Furdyna}\ \emph {et~al.}(2023)\citenamefont
  {Furdyna}, \citenamefont {Liu}, \citenamefont {Dobrowolska},\ and\
  \citenamefont {Lee}}]{Furdyna2023}%
  \BibitemOpen
  \bibfield  {author} {\bibinfo {author} {\bibfnamefont {J.~K.}\ \bibnamefont
  {Furdyna}}, \bibinfo {author} {\bibfnamefont {X.}~\bibnamefont {Liu}},
  \bibinfo {author} {\bibfnamefont {M.}~\bibnamefont {Dobrowolska}}, \ and\
  \bibinfo {author} {\bibfnamefont {S.}~\bibnamefont {Lee}},\ }\bibfield
  {title} {\enquote {\bibinfo {title} {Mn impurity band and the effects of {M}n
  position in {III}--{V} lattice: {P}ivotal contributions of {W}ladek
  {W}alukiewicz to the understanding of ferromagnetism in semiconductors},}\
  }\href {\doibase 10.1063/5.0176698} {\bibfield  {journal} {\bibinfo
  {journal} {J. Appl. Phys.}\ }\textbf {\bibinfo {volume} {134}},\ \bibinfo
  {pages} {200901} (\bibinfo {year} {2023})}\BibitemShut {NoStop}%
\bibitem [{\citenamefont {Berciu}\ and\ \citenamefont
  {Bhatt}(2001)}]{Berciu2001}%
  \BibitemOpen
  \bibfield  {author} {\bibinfo {author} {\bibfnamefont {M.}~\bibnamefont
  {Berciu}}\ and\ \bibinfo {author} {\bibfnamefont {R.~N.}\ \bibnamefont
  {Bhatt}},\ }\bibfield  {title} {\enquote {\bibinfo {title} {Effects of
  disorder on ferromagnetism in diluted magnetic semiconductors},}\ }\href
  {\doibase 10.1103/PhysRevLett.87.107203} {\bibfield  {journal} {\bibinfo
  {journal} {Phys. Rev. Lett.}\ }\textbf {\bibinfo {volume} {87}},\ \bibinfo
  {pages} {107203} (\bibinfo {year} {2001})}\BibitemShut {NoStop}%
\bibitem [{\citenamefont {Bhatt}\ \emph {et~al.}(2002)\citenamefont {Bhatt},
  \citenamefont {Berciu}, \citenamefont {Kennett},\ and\ \citenamefont
  {Wan}}]{Bhatt2002}%
  \BibitemOpen
  \bibfield  {author} {\bibinfo {author} {\bibfnamefont {R.~N.}\ \bibnamefont
  {Bhatt}}, \bibinfo {author} {\bibfnamefont {M.}~\bibnamefont {Berciu}},
  \bibinfo {author} {\bibfnamefont {M.~P.}\ \bibnamefont {Kennett}}, \ and\
  \bibinfo {author} {\bibfnamefont {X.}~\bibnamefont {Wan}},\ }\bibfield
  {title} {\enquote {\bibinfo {title} {Diluted magnetic semiconductors in the
  low carrier density regime},}\ }\href {\doibase
  https://dx.doi.org/10.1023/A:1014031327996} {\bibfield  {journal} {\bibinfo
  {journal} {J. Supercond. Nov. Magn.}\ }\textbf {\bibinfo {volume} {15}},\
  \bibinfo {pages} {71} (\bibinfo {year} {2002})}\BibitemShut {NoStop}%
\bibitem [{\citenamefont {Mayr}\ \emph {et~al.}(2002)\citenamefont {Mayr},
  \citenamefont {Alvarez},\ and\ \citenamefont {Dagotto}}]{Mayr2002}%
  \BibitemOpen
  \bibfield  {author} {\bibinfo {author} {\bibfnamefont {M.}~\bibnamefont
  {Mayr}}, \bibinfo {author} {\bibfnamefont {G.}~\bibnamefont {Alvarez}}, \
  and\ \bibinfo {author} {\bibfnamefont {E.}~\bibnamefont {Dagotto}},\
  }\bibfield  {title} {\enquote {\bibinfo {title} {Global versus local
  ferromagnetism in a model for diluted magnetic semiconductors studied with
  {M}onte {C}arlo techniques},}\ }\href {\doibase 10.1103/PhysRevB.65.241202}
  {\bibfield  {journal} {\bibinfo  {journal} {Phys. Rev. B}\ }\textbf {\bibinfo
  {volume} {65}},\ \bibinfo {pages} {241202} (\bibinfo {year}
  {2002})}\BibitemShut {NoStop}%
\bibitem [{\citenamefont {Kennett}\ \emph {et~al.}(2002)\citenamefont
  {Kennett}, \citenamefont {Berciu},\ and\ \citenamefont
  {Bhatt}}]{Kennett2002}%
  \BibitemOpen
  \bibfield  {author} {\bibinfo {author} {\bibfnamefont {M.~P.}\ \bibnamefont
  {Kennett}}, \bibinfo {author} {\bibfnamefont {M.}~\bibnamefont {Berciu}}, \
  and\ \bibinfo {author} {\bibfnamefont {R.~N.}\ \bibnamefont {Bhatt}},\
  }\bibfield  {title} {\enquote {\bibinfo {title} {Monte {C}arlo simulations of
  an impurity-band model for {III}--{V} diluted magnetic semiconductors},}\
  }\href {\doibase 10.1103/PhysRevB.66.045207} {\bibfield  {journal} {\bibinfo
  {journal} {Phys. Rev. B}\ }\textbf {\bibinfo {volume} {66}},\ \bibinfo
  {pages} {045207} (\bibinfo {year} {2002})}\BibitemShut {NoStop}%
\bibitem [{\citenamefont {Samarth}(2012)}]{Samarth2012}%
  \BibitemOpen
  \bibfield  {author} {\bibinfo {author} {\bibfnamefont {N.}~\bibnamefont
  {Samarth}},\ }\bibfield  {title} {\enquote {\bibinfo {title} {Battle of the
  bands},}\ }\href {https://www.nature.com/articles/nmat3317} {\bibfield
  {journal} {\bibinfo  {journal} {Nat. Mater.}\ }\textbf {\bibinfo {volume}
  {11}},\ \bibinfo {pages} {360} (\bibinfo {year} {2012})}\BibitemShut
  {NoStop}%
\bibitem [{\citenamefont {Yu}\ \emph {et~al.}(2008)\citenamefont {Yu},
  \citenamefont {Wojtowicz}, \citenamefont {Walukiewicz}, \citenamefont {Liu},\
  and\ \citenamefont {Furdyna}}]{Yu2008}%
  \BibitemOpen
  \bibfield  {author} {\bibinfo {author} {\bibfnamefont {K.M.}\ \bibnamefont
  {Yu}}, \bibinfo {author} {\bibfnamefont {T.}~\bibnamefont {Wojtowicz}},
  \bibinfo {author} {\bibfnamefont {W.}~\bibnamefont {Walukiewicz}}, \bibinfo
  {author} {\bibfnamefont {X.}~\bibnamefont {Liu}}, \ and\ \bibinfo {author}
  {\bibfnamefont {J.K.}\ \bibnamefont {Furdyna}},\ }\bibfield  {title}
  {\enquote {\bibinfo {title} {Chapter 3 {F}ermi level effects on {M}n
  incorporation in {III}--{M}n--{V} ferromagnetic semiconductors},}\ }in\ \href
  {\doibase https://doi.org/10.1016/S0080-8784(08)00003-3} {\emph {\bibinfo
  {booktitle} {Semiconductors and Semimetals}}},\ Vol.~\bibinfo {volume} {82}\
  (\bibinfo  {publisher} {Elsevier},\ \bibinfo {year} {2008})\ pp.\ \bibinfo
  {pages} {89--133}\BibitemShut {NoStop}%
\bibitem [{\citenamefont {Kokurin}\ \emph {et~al.}(2020)\citenamefont
  {Kokurin}, \citenamefont {Silov},\ and\ \citenamefont
  {Averkiev}}]{Kokurin2020}%
  \BibitemOpen
  \bibfield  {author} {\bibinfo {author} {\bibfnamefont {I.~A.}\ \bibnamefont
  {Kokurin}}, \bibinfo {author} {\bibfnamefont {A.~Yu.}\ \bibnamefont {Silov}},
  \ and\ \bibinfo {author} {\bibfnamefont {N.~S.}\ \bibnamefont {Averkiev}},\
  }\bibfield  {title} {\enquote {\bibinfo {title} {Sign-reversal electron
  magnetization in {M}n-doped semiconductor structures},}\ }\href {\doibase
  10.1103/PhysRevB.102.041202} {\bibfield  {journal} {\bibinfo  {journal}
  {Phys. Rev. B}\ }\textbf {\bibinfo {volume} {102}},\ \bibinfo {pages}
  {041202} (\bibinfo {year} {2020})}\BibitemShut {NoStop}%
\bibitem [{\citenamefont {Blinowski}\ and\ \citenamefont
  {Kacman}(2003)}]{Blinowski2003}%
  \BibitemOpen
  \bibfield  {author} {\bibinfo {author} {\bibfnamefont {J.}~\bibnamefont
  {Blinowski}}\ and\ \bibinfo {author} {\bibfnamefont {P.}~\bibnamefont
  {Kacman}},\ }\bibfield  {title} {\enquote {\bibinfo {title} {Spin
  interactions of interstitial {M}n ions in ferromagnetic {G}a{M}n{A}s},}\
  }\href {\doibase 10.1103/PhysRevB.67.121204} {\bibfield  {journal} {\bibinfo
  {journal} {Phys. Rev. B}\ }\textbf {\bibinfo {volume} {67}},\ \bibinfo
  {pages} {121204} (\bibinfo {year} {2003})}\BibitemShut {NoStop}%
\bibitem [{\citenamefont {Potashnik}\ \emph {et~al.}(2001)\citenamefont
  {Potashnik}, \citenamefont {Ku}, \citenamefont {Chun}, \citenamefont {Berry},
  \citenamefont {Samarth},\ and\ \citenamefont {Schiffer}}]{Potashnik2001}%
  \BibitemOpen
  \bibfield  {author} {\bibinfo {author} {\bibfnamefont {S.~J.}\ \bibnamefont
  {Potashnik}}, \bibinfo {author} {\bibfnamefont {K.~C.}\ \bibnamefont {Ku}},
  \bibinfo {author} {\bibfnamefont {S.~H.}\ \bibnamefont {Chun}}, \bibinfo
  {author} {\bibfnamefont {J.~J.}\ \bibnamefont {Berry}}, \bibinfo {author}
  {\bibfnamefont {N.}~\bibnamefont {Samarth}}, \ and\ \bibinfo {author}
  {\bibfnamefont {P.}~\bibnamefont {Schiffer}},\ }\bibfield  {title} {\enquote
  {\bibinfo {title} {Effects of annealing time on defect-controlled
  ferromagnetism in $\mathrm{Ga_{1-x}Mn_xAs}$},}\ }\href {\doibase
  10.1063/1.1398619} {\bibfield  {journal} {\bibinfo  {journal} {Appl. Phys.
  Lett.}\ }\textbf {\bibinfo {volume} {79}},\ \bibinfo {pages} {1495} (\bibinfo
  {year} {2001})}\BibitemShut {NoStop}%
\bibitem [{\citenamefont {Wojtowicz}\ \emph {et~al.}(2004)\citenamefont
  {Wojtowicz}, \citenamefont {Furdyna}, \citenamefont {Liu}, \citenamefont
  {Yu},\ and\ \citenamefont {Walukiewicz}}]{Wojtowicz2004}%
  \BibitemOpen
  \bibfield  {author} {\bibinfo {author} {\bibfnamefont {T.}~\bibnamefont
  {Wojtowicz}}, \bibinfo {author} {\bibfnamefont {J.K.}\ \bibnamefont
  {Furdyna}}, \bibinfo {author} {\bibfnamefont {X.}~\bibnamefont {Liu}},
  \bibinfo {author} {\bibfnamefont {K.M.}\ \bibnamefont {Yu}}, \ and\ \bibinfo
  {author} {\bibfnamefont {W.}~\bibnamefont {Walukiewicz}},\ }\bibfield
  {title} {\enquote {\bibinfo {title} {Electronic effects determining the
  formation of ferromagnetic $\mathrm{III_{1-x}Mn_xV}$ alloys during epitaxial
  growth},}\ }\href {\doibase https://doi.org/10.1016/j.physe.2004.06.014}
  {\bibfield  {journal} {\bibinfo  {journal} {Physica E (Amsterdam)}\ }\textbf
  {\bibinfo {volume} {25}},\ \bibinfo {pages} {171} (\bibinfo {year}
  {2004})}\BibitemShut {NoStop}%
\bibitem [{\citenamefont {Furdyna}\ \emph {et~al.}(2004)\citenamefont
  {Furdyna}, \citenamefont {Wojtowicz}, \citenamefont {Liu}, \citenamefont
  {Yu}, \citenamefont {Walukiewicz}, \citenamefont {Vurgaftman},\ and\
  \citenamefont {Meyer}}]{Furdyna2004}%
  \BibitemOpen
  \bibfield  {author} {\bibinfo {author} {\bibfnamefont {J.~K.}\ \bibnamefont
  {Furdyna}}, \bibinfo {author} {\bibfnamefont {T.}~\bibnamefont {Wojtowicz}},
  \bibinfo {author} {\bibfnamefont {X.}~\bibnamefont {Liu}}, \bibinfo {author}
  {\bibfnamefont {K.~M.}\ \bibnamefont {Yu}}, \bibinfo {author} {\bibfnamefont
  {W.}~\bibnamefont {Walukiewicz}}, \bibinfo {author} {\bibfnamefont
  {I.}~\bibnamefont {Vurgaftman}}, \ and\ \bibinfo {author} {\bibfnamefont
  {J.~R.}\ \bibnamefont {Meyer}},\ }\bibfield  {title} {\enquote {\bibinfo
  {title} {Fermi level effects on {M}n incorporation in modulation-doped
  ferromagnetic $\mathrm{III_{1-x}Mn_xV}$ heterostructures},}\ }\href {\doibase
  10.1088/0953-8984/16/48/004} {\bibfield  {journal} {\bibinfo  {journal} {J.
  Phys.: Condens. Matter}\ }\textbf {\bibinfo {volume} {16}},\ \bibinfo {pages}
  {S5499} (\bibinfo {year} {2004})}\BibitemShut {NoStop}%
\bibitem [{\citenamefont {Cho}\ \emph {et~al.}(2008{\natexlab{a}})\citenamefont
  {Cho}, \citenamefont {Yu}, \citenamefont {Liu}, \citenamefont {Walukiewicz},\
  and\ \citenamefont {Furdyna}}]{Cho2008a}%
  \BibitemOpen
  \bibfield  {author} {\bibinfo {author} {\bibfnamefont {Y.~J.}\ \bibnamefont
  {Cho}}, \bibinfo {author} {\bibfnamefont {K.~M.}\ \bibnamefont {Yu}},
  \bibinfo {author} {\bibfnamefont {X.}~\bibnamefont {Liu}}, \bibinfo {author}
  {\bibfnamefont {W.}~\bibnamefont {Walukiewicz}}, \ and\ \bibinfo {author}
  {\bibfnamefont {J.~K.}\ \bibnamefont {Furdyna}},\ }\bibfield  {title}
  {\enquote {\bibinfo {title} {Effects of donor doping on
  $\mathrm{Ga_{1-x}Mn_xAs}$},}\ }\href {\doibase 10.1063/1.3063046} {\bibfield
  {journal} {\bibinfo  {journal} {Appl. Phys. Lett.}\ }\textbf {\bibinfo
  {volume} {93}},\ \bibinfo {pages} {262505} (\bibinfo {year}
  {2008}{\natexlab{a}})}\BibitemShut {NoStop}%
\bibitem [{\citenamefont {Kokurin}\ and\ \citenamefont
  {Averkiev}(2024)}]{Kokurin2024}%
  \BibitemOpen
  \bibfield  {author} {\bibinfo {author} {\bibfnamefont {I.~A.}\ \bibnamefont
  {Kokurin}}\ and\ \bibinfo {author} {\bibfnamefont {N.~S.}\ \bibnamefont
  {Averkiev}},\ }\bibfield  {title} {\enquote {\bibinfo {title} {Model of
  localized state mediated exchange interaction and ferromagnetism in diluted
  magnetic semiconductors},}\ }\href {\doibase 10.1103/PhysRevB.109.214415}
  {\bibfield  {journal} {\bibinfo  {journal} {Phys. Rev. B}\ }\textbf {\bibinfo
  {volume} {109}},\ \bibinfo {pages} {214415} (\bibinfo {year}
  {2024})}\BibitemShut {NoStop}%
\bibitem [{\citenamefont {Kubo}(1988)}]{Kubo1988}%
  \BibitemOpen
  \bibfield  {author} {\bibinfo {author} {\bibfnamefont {R.}~\bibnamefont
  {Kubo}},\ }\href@noop {} {\emph {\bibinfo {title} {Statistical mechanics.
  {A}n Advanced Course with Problems and Solutions.}}},\ \bibinfo {edition}
  {7th}\ ed.\ (\bibinfo  {publisher} {North-Holland},\ \bibinfo {address}
  {Amsterdam},\ \bibinfo {year} {1988})\BibitemShut {NoStop}%
\bibitem [{\citenamefont {Averkiev}\ and\ \citenamefont
  {Il'inskii}(1994)}]{Averkiev1994}%
  \BibitemOpen
  \bibfield  {author} {\bibinfo {author} {\bibfnamefont {N.~S.}\ \bibnamefont
  {Averkiev}}\ and\ \bibinfo {author} {\bibfnamefont {S.~Yu.}\ \bibnamefont
  {Il'inskii}},\ }\bibfield  {title} {\enquote {\bibinfo {title} {Spin ordering
  of carriers localized at two deep centers in cubic semiconductors},}\
  }\href@noop {} {\bibfield  {journal} {\bibinfo  {journal} {Phys. Solid
  State}\ }\textbf {\bibinfo {volume} {36}},\ \bibinfo {pages} {278} (\bibinfo
  {year} {1994})}\BibitemShut {NoStop}%
\bibitem [{\citenamefont {Durst}\ \emph {et~al.}(2017)\citenamefont {Durst},
  \citenamefont {Castoria},\ and\ \citenamefont {Bhatt}}]{Durst2017}%
  \BibitemOpen
  \bibfield  {author} {\bibinfo {author} {\bibfnamefont {A.~C.}\ \bibnamefont
  {Durst}}, \bibinfo {author} {\bibfnamefont {K.~E.}\ \bibnamefont {Castoria}},
  \ and\ \bibinfo {author} {\bibfnamefont {R.~N.}\ \bibnamefont {Bhatt}},\
  }\bibfield  {title} {\enquote {\bibinfo {title} {Heitler-{L}ondon model for
  acceptor-acceptor interactions in doped semiconductors},}\ }\href {\doibase
  10.1103/PhysRevB.96.155208} {\bibfield  {journal} {\bibinfo  {journal} {Phys.
  Rev. B}\ }\textbf {\bibinfo {volume} {96}},\ \bibinfo {pages} {155208}
  (\bibinfo {year} {2017})}\BibitemShut {NoStop}%
\bibitem [{Note1()}]{Note1}%
  \BibitemOpen
  \bibinfo {note} {Formally, it is necessary to introduce independent
  magnetizations $\xi ^0$ and $\xi ^-$ to describe two charge states of
  manganese, ${\protect \rm Mn_{Ga}^0}$ and ${\protect \rm Mn_{Ga}^-}$,
  respectively. Due to the hole hopping, each Mn center regularly changes its
  charge state. Together with the mentioned negligible contribution of holes to
  the total magnetization, this allows one to describe the magnetization of both
  types of Mn centers using a single parameter $\xi $.}\BibitemShut {Stop}%
\bibitem [{Note2()}]{Note2}%
  \BibitemOpen
  \bibinfo {note} {The presence of ionized-ionized pairs is evident. However,
  these pairs do not contribute to the energy of indirect exchange interaction
  within the framework of our theory. The possible superexchange of nearest
  neighbors, however, can be taken into account simply by renormalization of
  AFM pair contribution for the pairs being in the first CS.}\BibitemShut
  {Stop}%
\bibitem [{\citenamefont {Thomas}\ \emph {et~al.}(1964)\citenamefont {Thomas},
  \citenamefont {Gershenzon},\ and\ \citenamefont {Trumbore}}]{Thomas1964}%
  \BibitemOpen
  \bibfield  {author} {\bibinfo {author} {\bibfnamefont {D.~G.}\ \bibnamefont
  {Thomas}}, \bibinfo {author} {\bibfnamefont {M.}~\bibnamefont {Gershenzon}},
  \ and\ \bibinfo {author} {\bibfnamefont {F.~A.}\ \bibnamefont {Trumbore}},\
  }\bibfield  {title} {\enquote {\bibinfo {title} {Pair spectra and "edge"
  emission in gallium phosphide},}\ }\href {\doibase 10.1103/PhysRev.133.A269}
  {\bibfield  {journal} {\bibinfo  {journal} {Phys. Rev.}\ }\textbf {\bibinfo
  {volume} {133}},\ \bibinfo {pages} {A269} (\bibinfo {year}
  {1964})}\BibitemShut {NoStop}%
\bibitem [{Note3()}]{Note3}%
  \BibitemOpen
  \bibinfo {note} {It should be remembered that by the same method (as for
  ionized-neutral pairs) each neutral-neutral pair is counted
  twice.}\BibitemShut {Stop}%
\bibitem [{Note4()}]{Note4}%
  \BibitemOpen
  \bibinfo {note} {In diluted systems the number of possible `magnetic'
  configurations should be multiplied by number of possible dilution
  configurations $P(x)=\protect \frac {N!}{(xN)![(1-x)N]!}$. This leads to the
  appearance of an additional magnetization-independent term in the free energy
  $F$. This contribution can be omitted when calculating the magnetic
  properties.}\BibitemShut {Stop}%
\bibitem [{Note5()}]{Note5}%
  \BibitemOpen
  \bibinfo {note} {Here the Stirling's approximation is used for $S=k_B\protect
  \qopname \relax o{ln}P^m(\xi )$ calculation, $\protect \qopname \relax
  o{ln}n!=n\protect \qopname \relax o{ln}n - n$. In Eq.~(\ref {free_energy})
  $\xi $-independent term $-xNk_BT\protect \qopname \relax o{ln}2$ is
  omitted.}\BibitemShut {Stop}%
\bibitem [{\citenamefont {Smart}(1966)}]{Smart1966}%
  \BibitemOpen
  \bibfield  {author} {\bibinfo {author} {\bibfnamefont {J.S.}\ \bibnamefont
  {Smart}},\ }\href@noop {} {\emph {\bibinfo {title} {Effective Field Theories
  of Magnetism}}}\ (\bibinfo  {publisher} {Saunders},\ \bibinfo {address}
  {Philadelphia},\ \bibinfo {year} {1966})\BibitemShut {NoStop}%
\bibitem [{\citenamefont {Williams}(1968)}]{Williams1968}%
  \BibitemOpen
  \bibfield  {author} {\bibinfo {author} {\bibfnamefont {F.}~\bibnamefont
  {Williams}},\ }\bibfield  {title} {\enquote {\bibinfo {title} {Donor--acceptor
  pairs in semiconductors},}\ }\href {\doibase
  https://doi.org/10.1002/pssb.19680250202} {\bibfield  {journal} {\bibinfo
  {journal} {Phys. Status Solidi}\ }\textbf {\bibinfo {volume} {25}},\ \bibinfo
  {pages} {493} (\bibinfo {year} {1968})}\BibitemShut {NoStop}%
\bibitem [{Note6()}]{Note6}%
  \BibitemOpen
  \bibinfo {note} {The ionization energy of ${\protect \rm Mn_{Ga}^0}$ is about
  110 meV \cite {Averkiev2018}, that corresponds to level with total angular
  momentum $F=1$. The `bare' (without exchange) ionization energy is close to
  $E_0=96.9$ meV, when the exchange parameter $A$ is equal to $-2.5$ meV \cite
  {Sapega2001}. Thus $A/E_0$=-0.026. For real ${\protect \rm Mn_{Ga}^0}$
  acceptor the energy distance between FM- and AFM-aligned states ($F=4$ and
  $F=1$) is equal to $9|A|$, i.e., relative magnitude $9|A|/E_0$ is 0.232. This
  justifies the use of the value $A/E_0=-0.2$ in the case of a simplified
  1/2-spin model.}\BibitemShut {Stop}%
\bibitem [{\citenamefont {Liu}\ \emph {et~al.}(1995)\citenamefont {Liu},
  \citenamefont {Prasad}, \citenamefont {Nishio}, \citenamefont {Weber},
  \citenamefont {Liliental-Weber},\ and\ \citenamefont
  {Walukiewicz}}]{Liu1995}%
  \BibitemOpen
  \bibfield  {author} {\bibinfo {author} {\bibfnamefont {X.}~\bibnamefont
  {Liu}}, \bibinfo {author} {\bibfnamefont {A.}~\bibnamefont {Prasad}},
  \bibinfo {author} {\bibfnamefont {J.}~\bibnamefont {Nishio}}, \bibinfo
  {author} {\bibfnamefont {E.~R.}\ \bibnamefont {Weber}}, \bibinfo {author}
  {\bibfnamefont {Z.}~\bibnamefont {Liliental-Weber}}, \ and\ \bibinfo {author}
  {\bibfnamefont {W.}~\bibnamefont {Walukiewicz}},\ }\bibfield  {title}
  {\enquote {\bibinfo {title} {Native point defects in low-temperature-grown
  {G}a{A}s},}\ }\href {\doibase 10.1063/1.114782} {\bibfield  {journal}
  {\bibinfo  {journal} {Appl. Phys. Lett.}\ }\textbf {\bibinfo {volume} {67}},\
  \bibinfo {pages} {279} (\bibinfo {year} {1995})}\BibitemShut {NoStop}%
\bibitem [{\citenamefont {Grandidier}\ \emph {et~al.}(2000)\citenamefont
  {Grandidier}, \citenamefont {Nys}, \citenamefont {Delerue}, \citenamefont
  {Stievenard}, \citenamefont {Higo},\ and\ \citenamefont
  {Tanaka}}]{Grandidier2000}%
  \BibitemOpen
  \bibfield  {author} {\bibinfo {author} {\bibfnamefont {B.}~\bibnamefont
  {Grandidier}}, \bibinfo {author} {\bibfnamefont {J.~P.}\ \bibnamefont {Nys}},
  \bibinfo {author} {\bibfnamefont {C.}~\bibnamefont {Delerue}}, \bibinfo
  {author} {\bibfnamefont {D.}~\bibnamefont {Stievenard}}, \bibinfo {author}
  {\bibfnamefont {Y.}~\bibnamefont {Higo}}, \ and\ \bibinfo {author}
  {\bibfnamefont {M.}~\bibnamefont {Tanaka}},\ }\bibfield  {title} {\enquote
  {\bibinfo {title} {Atomic-scale study of {G}a{M}n{A}s/{G}a{A}s layers},}\
  }\href {\doibase 10.1063/1.1322052} {\bibfield  {journal} {\bibinfo
  {journal} {Appl. Phys. Lett.}\ }\textbf {\bibinfo {volume} {77}},\ \bibinfo
  {pages} {4001} (\bibinfo {year} {2000})}\BibitemShut {NoStop}%
\bibitem [{\citenamefont {Kokurin}\ and\ \citenamefont
  {Averkiev}(2023)}]{Kokurin2023}%
  \BibitemOpen
  \bibfield  {author} {\bibinfo {author} {\bibfnamefont {I.~A.}\ \bibnamefont
  {Kokurin}}\ and\ \bibinfo {author} {\bibfnamefont {N.~S.}\ \bibnamefont
  {Averkiev}},\ }\bibfield  {title} {\enquote {\bibinfo {title} {Level
  structure of hole two-center complex and related luminescence in
  semiconductors},}\ }\href {\doibase 10.1103/PhysRevB.107.125208} {\bibfield
  {journal} {\bibinfo  {journal} {Phys. Rev. B}\ }\textbf {\bibinfo {volume}
  {107}},\ \bibinfo {pages} {125208} (\bibinfo {year} {2023})}\BibitemShut
  {NoStop}%
\bibitem [{\citenamefont {Karpenko}\ and\ \citenamefont
  {Berdyshev}(1964)}]{Karpenko1964}%
  \BibitemOpen
  \bibfield  {author} {\bibinfo {author} {\bibfnamefont {B.~V.}\ \bibnamefont
  {Karpenko}}\ and\ \bibinfo {author} {\bibfnamefont {A.~A.}\ \bibnamefont
  {Berdyshev}},\ }\bibfield  {title} {\enquote {\bibinfo {title} {Exchange
  interaction via current carriers in ordered semiconducting magnets},}\
  }\href@noop {} {\bibfield  {journal} {\bibinfo  {journal} {Sov. Phys. Solid
  State}\ }\textbf {\bibinfo {volume} {5}},\ \bibinfo {pages} {2494} (\bibinfo
  {year} {1964})}\BibitemShut {NoStop}%
\bibitem [{\citenamefont {Berdyshev}(1966)}]{Berdyshev1966}%
  \BibitemOpen
  \bibfield  {author} {\bibinfo {author} {\bibfnamefont {A.~A.}\ \bibnamefont
  {Berdyshev}},\ }\bibfield  {title} {\enquote {\bibinfo {title} {Ferromagnetic
  semiconductors in which the exchange coupling occurs through the conduction
  electrons},}\ }\href@noop {} {\bibfield  {journal} {\bibinfo  {journal} {Sov.
  Phys. Solid State}\ }\textbf {\bibinfo {volume} {8}},\ \bibinfo {pages}
  {1104} (\bibinfo {year} {1966})}\BibitemShut {NoStop}%
\bibitem [{\citenamefont {Chapman}\ and\ \citenamefont
  {Hutchinson}(1967)}]{Chapman1967}%
  \BibitemOpen
  \bibfield  {author} {\bibinfo {author} {\bibfnamefont {R.~A.}\ \bibnamefont
  {Chapman}}\ and\ \bibinfo {author} {\bibfnamefont {W.~G.}\ \bibnamefont
  {Hutchinson}},\ }\bibfield  {title} {\enquote {\bibinfo {title}
  {Photoexcitation and photoionization of neutral manganese acceptors in
  gallium arsenide},}\ }\href {\doibase 10.1103/PhysRevLett.18.443} {\bibfield
  {journal} {\bibinfo  {journal} {Phys. Rev. Lett.}\ }\textbf {\bibinfo
  {volume} {18}},\ \bibinfo {pages} {443} (\bibinfo {year} {1967})}\BibitemShut
  {NoStop}%
\bibitem [{\citenamefont {Brown}\ \emph {et~al.}(1973)\citenamefont {Brown},
  \citenamefont {Woodbury},\ and\ \citenamefont {Blakemore}}]{Brown1973}%
  \BibitemOpen
  \bibfield  {author} {\bibinfo {author} {\bibfnamefont {W.~J.}\ \bibnamefont
  {Brown}}, \bibinfo {author} {\bibfnamefont {D.~A.}\ \bibnamefont {Woodbury}},
  \ and\ \bibinfo {author} {\bibfnamefont {J.~S.}\ \bibnamefont {Blakemore}},\
  }\bibfield  {title} {\enquote {\bibinfo {title} {Photoionization cross
  section for manganese acceptors in gallium arsenide},}\ }\href {\doibase
  10.1103/PhysRevB.8.5664} {\bibfield  {journal} {\bibinfo  {journal} {Phys.
  Rev. B}\ }\textbf {\bibinfo {volume} {8}},\ \bibinfo {pages} {5664} (\bibinfo
  {year} {1973})}\BibitemShut {NoStop}%
\bibitem [{\citenamefont {Kokurin}\ \emph {et~al.}(2013)\citenamefont
  {Kokurin}, \citenamefont {Petrov},\ and\ \citenamefont
  {Averkiev}}]{Kokurin2013}%
  \BibitemOpen
  \bibfield  {author} {\bibinfo {author} {\bibfnamefont {I.A.}\ \bibnamefont
  {Kokurin}}, \bibinfo {author} {\bibfnamefont {P.V.}\ \bibnamefont {Petrov}},
  \ and\ \bibinfo {author} {\bibfnamefont {N.S.}\ \bibnamefont {Averkiev}},\
  }\bibfield  {title} {\enquote {\bibinfo {title} {Optical orientation of
  electrons in compensated semiconductors},}\ }\href {\doibase
  10.1134/S106378261309011X} {\bibfield  {journal} {\bibinfo  {journal}
  {Semiconductors}\ }\textbf {\bibinfo {volume} {47}},\ \bibinfo {pages} {1232}
  (\bibinfo {year} {2013})}\BibitemShut {NoStop}%
\bibitem [{\citenamefont {Petrov}\ \emph {et~al.}(2016)\citenamefont {Petrov},
  \citenamefont {Kokurin}, \citenamefont {Ivanov}, \citenamefont {Averkiev},
  \citenamefont {Campion}, \citenamefont {Gallagher}, \citenamefont
  {Koenraad},\ and\ \citenamefont {Silov}}]{Petrov2016}%
  \BibitemOpen
  \bibfield  {author} {\bibinfo {author} {\bibfnamefont {P.~V.}\ \bibnamefont
  {Petrov}}, \bibinfo {author} {\bibfnamefont {I.~A.}\ \bibnamefont {Kokurin}},
  \bibinfo {author} {\bibfnamefont {Yu.~L.}\ \bibnamefont {Ivanov}}, \bibinfo
  {author} {\bibfnamefont {N.~S.}\ \bibnamefont {Averkiev}}, \bibinfo {author}
  {\bibfnamefont {R.~P.}\ \bibnamefont {Campion}}, \bibinfo {author}
  {\bibfnamefont {B.~L.}\ \bibnamefont {Gallagher}}, \bibinfo {author}
  {\bibfnamefont {P.~M.}\ \bibnamefont {Koenraad}}, \ and\ \bibinfo {author}
  {\bibfnamefont {A.~Yu.}\ \bibnamefont {Silov}},\ }\bibfield  {title}
  {\enquote {\bibinfo {title} {Optical orientation of spins in
  {G}a{A}s:{M}n/{A}l{G}a{A}s quantum wells via impurity-to-band excitation},}\
  }\href {\doibase 10.1103/PhysRevB.94.085308} {\bibfield  {journal} {\bibinfo
  {journal} {Phys. Rev. B}\ }\textbf {\bibinfo {volume} {94}},\ \bibinfo
  {pages} {085308} (\bibinfo {year} {2016})}\BibitemShut {NoStop}%
\bibitem [{\citenamefont {Cho}\ \emph {et~al.}(2008{\natexlab{b}})\citenamefont
  {Cho}, \citenamefont {Liu},\ and\ \citenamefont {Furdyna}}]{Cho2008}%
  \BibitemOpen
  \bibfield  {author} {\bibinfo {author} {\bibfnamefont {Y.~J.}\ \bibnamefont
  {Cho}}, \bibinfo {author} {\bibfnamefont {X.}~\bibnamefont {Liu}}, \ and\
  \bibinfo {author} {\bibfnamefont {J.~K.}\ \bibnamefont {Furdyna}},\
  }\bibfield  {title} {\enquote {\bibinfo {title} {Collapse of ferromagnetism
  in ({G}a,{M}n){A}s at high hole concentrations},}\ }\href {\doibase
  10.1088/0268-1242/23/12/125010} {\bibfield  {journal} {\bibinfo  {journal}
  {Semicond. Sci. Technol.}\ }\textbf {\bibinfo {volume} {23}},\ \bibinfo
  {pages} {125010} (\bibinfo {year} {2008}{\natexlab{b}})}\BibitemShut
  {NoStop}%
\bibitem [{Note7()}]{Note7}%
  \BibitemOpen
  \bibinfo {note} {The average value $J^\mathrm{FM}=2$ meV at short intercenter
  distances together with $a=1.12$ eV leads to $\DOTSB \sum@ \slimits@
  _{j=1}^{j_{max}}Z_j=560$, i.e. $j_{max}=20$, which corresponds to
  $R_j=3.16a_0$.}\BibitemShut {Stop}%
\bibitem [{\citenamefont {Averkiev}\ and\ \citenamefont
  {Gutkin}(2018)}]{Averkiev2018}%
  \BibitemOpen
  \bibfield  {author} {\bibinfo {author} {\bibfnamefont {N.~S.}\ \bibnamefont
  {Averkiev}}\ and\ \bibinfo {author} {\bibfnamefont {A.~A.}\ \bibnamefont
  {Gutkin}},\ }\bibfield  {title} {\enquote {\bibinfo {title}
  {$\mathrm{Mn_{Ga}}$ acceptor center in {G}a{A}s ({R}eview)},}\ }\href
  {https://doi.org/10.1134/S106378341812003X} {\bibfield  {journal} {\bibinfo
  {journal} {Phys. Solid State}\ }\textbf {\bibinfo {volume} {60}},\ \bibinfo
  {pages} {2311} (\bibinfo {year} {2018})}\BibitemShut {NoStop}%
\bibitem [{\citenamefont {Sapega}\ \emph {et~al.}(2001)\citenamefont {Sapega},
  \citenamefont {Ruf},\ and\ \citenamefont {Cardona}}]{Sapega2001}%
  \BibitemOpen
  \bibfield  {author} {\bibinfo {author} {\bibfnamefont {V.F.}\ \bibnamefont
  {Sapega}}, \bibinfo {author} {\bibfnamefont {T.}~\bibnamefont {Ruf}}, \ and\
  \bibinfo {author} {\bibfnamefont {M.}~\bibnamefont {Cardona}},\ }\bibfield
  {title} {\enquote {\bibinfo {title} {Spin-flip Raman study of exchange
  interactions in bulk {G}a{A}s:{M}n},}\ }\href {\doibase
  10.1002/1521-3951(200108)226:2<339::AID-PSSB339>3.0.CO;2-8} {\bibfield
  {journal} {\bibinfo  {journal} {Phys. Status Solidi B}\ }\textbf {\bibinfo
  {volume} {226}},\ \bibinfo {pages} {339} (\bibinfo {year}
  {2001})}\BibitemShut {NoStop}%

\end{thebibliography}

%

\end{document}